\documentclass[fp,twocolumn]{jpsj3}
\usepackage{txfonts}
\usepackage{color}
\usepackage{bm}

\title{Nanorod Size Dependence of Coherent Coupling between Individual and Collective Excitations via Transverse Electromagnetic Field}

\author{Masayuki Iio$^{1}$, Tomohiro Yokoyama$^2$\thanks{{tomohiro.yokoyama@mp.es.}osaka-u.ac.jp}, Takeshi Inaoka$^3$, and Hajime Ishihara$^{2,4}$}
\inst{$^1$Department of Physics and Electronics, Graduate School of Engineering,
Osaka Prefecture University, 1-1 Gakuen-cho, Naka-ku, Sakai, Osaka 599-8531, Japan \\
$^2$Department of Materials Engineering Science, Graduate School of Engineering Science,
Osaka University, 1-3 Machikaneyama, Toyonaka, Osaka 560-8531, Japan\\
$^3$Department of Physics and Earth Sciences, Faculty of Science,
University of the Ryukyus, 1 Senbaru, Nishihara-cho, Okinawa 903-0213, Japan \\
$^4$Center for Quantum Information and Quantum Biology, Osaka University, 1-3 Machikaneyama, Toyonaka, Osaka 560-8531, Japan} %\\

\abst{
Plasmon is a collective excitation in metals formed by the Coulomb interaction between individual excitations of electron-hole pairs.
In many previous studies on the plasmonic response, the role of the longitudinal field has been almost exclusively focused on
light-induced plasmonic phenomena, such as hot carrier generation.
In our previous study [Phys.\ Rev.\ B \textbf{105}, 165408 (2022)], we have demonstrated the significant contribution of
the transverse electromagnetic field in connecting plasmons and electron-hole pairs in nanostructures based on
the self-consistent and nonlocal response theory. 
In this study, we investigate how this contribution appears depending on the system parameters, such as, rod length and refractive index.
The elucidation of the role of coherent coupling between the collective and individual excitations by the transverse field
will lead to the principle of controlling the bidirectional energy transfer between the plasmons and the electron-hole pairs,
which could significantly improve the efficiency of the hot carrier generation.
}

\begin{document}
\maketitle

\section{Introduction}
Plasmon is one of the fundamental collective excitation modes of electrons.
The excitation and relaxation mechanisms of the plasmons in metallic structures by
a light irradiation have been investigated extensively based on phenomenological relaxation processes~\cite{Brongersma15,Besteiro17}.
The plasmon excited by the light can relax non-radiatively in addition to the radiative light emission process.
The nonradiative relaxation causes a thermally non-equilibrium distribution of electrons and holes in ten femtosecond timescale.
They are called hot carriers (HCs).
The non-equilibrium distribution relaxes to a high temperature thermal distribution in picosecond timescale.
This light-induced HC generation and injection~\cite{Brongersma15,Besteiro17,White12,Govorov13,Govorov14,Kumar19}
have attracted considerable attention for applications, such as photocatalysis~\cite{Ueno16}, photodetection~\cite{Li17},
photocarrier injection~\cite{Tatsuma17}, and photovoltaics~\cite{Clavero14},
because the HCs have higher energy than the Schottky barrier between hosting metal and attached semiconductor~\cite{Goykhman11,White12}.

The Coulomb interaction between the individual electron-hole pair excitations forms a plasmon excitation.
The Coulomb interaction is mediated by the longitudinal (L) component of the electric field.
In bulk, the states of the collective excitation are orthogonal to each of the individual excitations.
However, in nanostructures, they are hybridized with each other, namely, the quantum coherence exists between
the collective and individual electron-hole pair excitations, where the size effect for electrons becomes significant.
Therefore, the HC generation depends on the size and shape of the materials hosting the plasmons~\cite{Govorov13,Govorov14},
and a microscopic approach is necessary to understand the HC generations from plasmon.
To describe such coupling based on a quantum microscopic treatment, for example, 
the first-principle calculations have been applied to nanoscale clusters~\cite{Ma15,You18}.

On surfaces and in nanostructures, the plasmon is excited by an incident light.
The light is the transverse (T) component of the electromagnetic field.
Moreover, photons could be re-emitted from the plasmon during the relaxation process.
The T field follows the Maxwell's equations.
Hence it should be considered self-consistently with both the plasmonic collective and the individual electron-hole excitations.
Furthermore, substrate fabrication to capture the incident light can increase the HC generation efficiency~\cite{XShi18,YELiu23}.
Hence, how the spatial structure of the electromagnetic field, including the T component,
affects the efficiency of plasmonic devices is an important subject.
An approach for self-consistent treatment of plasmon excitation and electromagnetic fields has been examined by using
the hydrodynamic Drude model~\cite{Bennett70,Schwartz82,Pitarke07,Mortensen14,Christensen14,Svendsen20}.
The hydrodynamic model treats a spatial gradient of current density in a conventional Drude conductivity,
which describes phenomenologically a nonlocal response.
However, the problem of how the nonlocal and self-consistent interaction among 
the plasmon and electron-hole pairs excitations and the T component of the Maxwell field has been poorly studied thus far.

In our recent study~\cite{Yokoyama22}, we have formulated a microscopic nonlocal theory for nanoscale electronic systems
and have revealed that the T field makes a finite contribution to the coupling between the plasmons and electron-hole pairs.
In our formulation, the constitutive equation with the nonlocal susceptibility and the Maxwell's equations with Green's function
are considered self-consistently~\cite{book:cho1}.
The electromagnetic fields are described as the vector and scalar potentials, ($\bm{A}, \phi$),
representing the T and L components, respectively, in the Coulomb gauge,
whereas the induced excitations are described in terms of current and charge densities, ($\bm{j}, \rho$).
The nonlocal susceptibility is deduced from the microscopic Hamiltonian and the linear response theory~\cite{Kubo57}.
Its nonlocality of the electronic response is attributed to the spatial extension of the electron wavefunctions.
%Therefore, the microscopic understanding is essential for an enhancement of light--matter interaction in nanostructures.
Based on the self-consistent formulation, we have examined the coherent coupling between
the collective and individual excitations in a single nanorod and
have found a finite contribution of the T field on the plasmon energy to the coherent coupling.
The coherent coupling enables a significant bidirectional energy transfer between the collective and individual excitations
when the nanostructure was designed properly.
The background dielectric constant of the nanorod modulates the wavelength of the T field.
Then, the coherent coupling due to the T field might be enlarged and optimized by the choice of size, shape, material, etc.

It is important to discuss to
%what extent the T-field-mediated coupling prominently appears in the realistic systems.
the extent to which the T-field-mediated coupling is prominent in the realistic systems. 
For the prominent appearance of such a coupling, the sufficient matching between the T field and electronic wavefunctions is necessary.
However, for the metals, the number of electrons contributing to the formation of collective excitation is considerably large in realistic systems,
and numerical examination is not easy. 
Therefore, as a first step, the aim of this study is to examine how the effect appears depending on the system parameters by
using a small set of electronic systems.
Although the demonstrated effect is not prominent, the obtained behaviors against the system parameters
have essential information to foresee the extent to which an effect appears in a realistic system.

For a quantitative discussion, we examine parameter modulation.
In our formulation, we can continuously tune the contribution of the T field components.
Then, the bidirectional energy transfer owing to the coherent coupling by the T field is
evaluated as a shift of the collective excitation energy during the tuning.
The coherent coupling owing to the T field depends on the nanorod length $L_z$ and the background refractive index $n_{\rm b}$.
Here, $n_{\rm b}$ causes two effects: modulation of the wavelength and screening of the Coulomb interaction.
The light-matter interaction could be enhanced owing to these effects and the length $L_z$ by
matching the spatial structures of the electronic wavefunctions and the T field.
When $n_{\rm b}$ is increased, the collective excitation energy decreases by the screening effect,
namely, the wavelength corresponding to the plasmon energy becomes slightly longer with the increase of $n_{\rm b}$.
On the other hand, the T field wavelength is shortened with the increase of $n_{\rm b}$.
How these counter effects appear as the energy shift of collective excitation should be examined.
Furthermore, we examine the influence of the Fermi energy of electrons in the nanorod.
Although the effect of the T field is not prominent for our small electronic bases set, 
these examinations reveal the minimum system size at which the finite effect of the T field begins to appear significantly. 
This study will suggest the possibility of spontaneous resonance by the coupling of plasmon and electron-hole pair excitations
that leads to the efficient HC generation, which should be demonstrated by large scale computations in the next step.  

The remainder of this article is structured as follows.
In Sec.\ \ref{sec:model}, we describe the microscopic Hamiltonian and a self-consistent formulation.
The formulation is applied to numerical calculations in Sec.\ \ref{sec:result}.
In the numrical calculation, we consider a single nanorod with various refractive indices and nanorod lengths.
Then, Sec.\ \ref{sec:conculusion} is devoted to the conclusions of this study.

\section{Model and Method}
\label{sec:model}
In this section, we describe a model to examine the coherent coupling between
the collective and individual excitations in nanostructures.
The formulation follows our previous study~\cite{Yokoyama22}.
The self-consistent treatment of the constitutive and Maxwell's equations provides a matrix equation.
Denoting the matrix as $\bar{\Xi} (\omega)$, the complex roots of ${\rm det} [\bar{\Xi} (\omega)] = 0$ provide
the individual and collective excitation energies (real components) with radiative damping (imaginary components).

\subsection{Hamiltonian for the light--matter interaction}
The Hamiltonian for electrons in an electromagnetic field is described as
\begin{align}
\hat{H} &= \sum_j 
\frac{\left\{ \hat{\bm{p}}_j - e \int d\bm{r} \bm{A} (\bm{r},t) \delta (\bm{r} - \hat{\bm{r}}_j) \right\}^2}{2m^*}
%- \varepsilon_{\rm F}
\nonumber \\
& + e \sum_j \int d\bm{r} \phi (\bm{r},t) \delta (\bm{r} - \hat{\bm{r}}_j),
\nonumber \\
& + \frac{1}{2} \sum_{i \ne j} \frac{e^2}{4\pi \epsilon_0 |\hat{\bm{r}}_i - \hat{\bm{r}}_j|},
\label{eq:singleH}
\end{align}
where $m^*$ and $e$ are the electron effective mass and charge, respectively, and $\epsilon_0$ is the dielectric constant of the vacuum.
%$\varepsilon_{\rm F}$ is the Fermi energy in the nanostructure.
For the Coulomb gauge, the vector potential satisfies ${\rm div} \bm{A} = 0$ and describes the T field.
The scalar potential gives the L field, which is divided into the nuclei and external origins,
$\phi (\bm{r},t) = \phi_{\rm ncl} ({\bm{r}}) + \phi_{\rm ext} (\bm{r},t)$.
%%% for reply
We assume that the nucleis are fixed.
Note that the Coulomb interaction in the third therm in Eq.\ (\ref{eq:singleH}) is a part of the L component.

We consider the second quantization for the Hamiltonian $\hat{H}$.
In the absence of external fields, $\bm{A} =0$ and $\phi_{\rm ext} =0$, we obtain
\begin{align}
\hat{H}_0 &= \sum_{n,n^\prime} {\psi_{n^\prime}^* (\bm{x})
\left( -\frac{\hbar^2}{2m^*} \bm{\nabla}_x^2 \right)
\psi_n (\bm{x}) \hat{a}_{n^\prime}^\dagger} \hat{a}_n
\nonumber \\
& + \int d\bm{x} \sum_{n,n^\prime} \hat{a}_{n^\prime}^\dagger {\psi_{n^\prime}^* (\bm{x})}
\left[ \phi_{\rm ncl} ({\bm{x}}) + \hat{\phi}_{\rm e-e} ({\bm{x}}) \right] {\psi_n (\bm{x})} \hat{a}_n,
\label{eq:H0}
\end{align}
with
\begin{equation}
\hat{\phi}_{\rm e-e} ({\bm{x}}) = \int d\bm{x}^\prime
{e \sum_{m, m^\prime} \frac{\psi_{m^\prime}^* (\bm{x}^\prime) \psi_m (\bm{x}^\prime)}{4\pi \epsilon_0 |\bm{x} - \bm{x}^\prime|}
 \hat{a}_{m^\prime}^\dagger \hat{a}_m,}
\label{eq:eescalar}
\end{equation}
describing the ``inner'' Coulomb interaction.
Here, $\hat{a}_n^\dagger$ and $\hat{a}_n$ are the creation and annihilation operators of the electrons of the state $n$, respectively.
$\psi_n (\bm{x})$ is the wavefunction of the state $n$.
We treat $\hat{H}_0$ as a nonperturbative Hamiltonian, which has no time dependence. 
Note that for an electrically neutral of material, $\langle \phi_{\rm ncl} + \hat{\phi}_{\rm e-e} \rangle_0 = 0$ must be satisfied.
Here, $\langle \cdots \rangle_0$ denotes the statistical average of the system without an external field.
The electron density in a static situation is given as
$\rho_0 (\bm{x})/e = \sum_n \psi_n^* (\bm{x}) \psi_n (\bm{x}) \langle \hat{a}_n^\dagger \hat{a}_n \rangle_0$.
Hence, the charge nutral condition is rewritten as
$\phi_{\rm ncl} (\bm{x}) = - \int d\bm{x}^\prime \frac{\rho_0 (\bm{x}^\prime)}{4 \pi \epsilon_0 |\bm{x} - \bm{x}^\prime|}$.

Let us consider a perturbative Hamiltonian $\hat{H}^\prime$ when an external field is applied,
\begin{equation}
\hat{H} = \hat{H}_0 + \hat{H}^\prime.
\label{eq:fullH}
\end{equation}
We assume a monochromatic electromagnetic field, $\bm{A} (\bm{x},t) = \bm{A} (\bm{x};\omega) e^{-i\omega t}$.
Then,
\begin{equation}
\hat{H}^\prime \simeq - \int d\bm{x} \left[
\hat{\bm{j}} {(\bm{x})} \cdot \bm{A} (\bm{x};\omega) - \delta \hat{\rho} {(\bm{x})} \phi_{\rm ext} (\bm{x};\omega)
%\Big( \phi_{\rm ext} (\bm{x},t) + \hat{\phi}_{\rm pol} (\bm{x},t) \Big)
\right] e^{-i\omega t}.
\label{eq:Hprime}
\end{equation}
Here, $\hat{\bm{j}} (\bm{x})$ and $\delta \hat{\rho} (\bm{x}) = \hat{\rho} (\bm{x}) - \rho_0 (\bm{x})$ are
the induced current and the (deviation of) charge density operators, respectively,
\begin{align}
\hat{\bm{j}} (\bm{x}) &= \sum_{n,n^\prime} \bm{j}_{n^\prime n} (\bm{x}) \hat{a}_{n^\prime}^\dagger \hat{a}_n, \\
%%%
\delta \hat{\rho} (\bm{x}) &= \sum_{n,n^\prime} \rho_{n^\prime n} (\bm{x}) \hat{a}_{n^\prime}^\dagger \hat{a}_n - \rho_0 (\bm{x}).
\end{align}
%$\delta \hat{\rho} (\bm{x}) = \hat{\rho} (\bm{x}) - \rho_0 (\bm{x})$ means the deviation of charge from the nutral.
The matrix elements for the current and charge are
\begin{align}
\rho_{n^\prime n} (\bm{x}) &= e \psi_{n^\prime}^* (\bm{x}) \psi_n (\bm{x}) \\
\bm{j}_{n^\prime n} (\bm{x}) &= - \frac{e\hbar}{2im^*} \Big[ \left\{ \bm{\nabla}_x \psi_{n^\prime}^* (\bm{x}) \right\} \psi_n (\bm{x})
- \psi_{n^\prime}^* (\bm{x}) \left\{ \bm{\nabla}_x \psi_n (\bm{x}) \right\} \Big]
\nonumber \\
& - \frac{e^2}{m^*} \psi_{n^\prime}^* (\bm{x}) \bm{A} (\bm{x};\omega) \psi_n (\bm{x}).
\end{align}
%with $\psi_n (\bm{x})$ is the wavefunction of state $n$.
In the following, we consider a self-consistent treatment of the induced densities and the external fields with the Maxwell's equations.
In the second term in Eq.\ (\ref{eq:Hprime}), the interaction between the external field and the positive background charge due to the nuclei is included.
For the self-consistent treatment, the interaction between the induced charge densities should be obtained.
We consider an additional scalar potential
\begin{equation}
\hat{\phi}_{\rm pol} (\bm{x}) = \int d\bm{x}^\prime
\frac{\delta \hat{\rho} (\bm{x}^\prime)}{4\pi \epsilon_0 |\bm{x} - \bm{x}^\prime|}
\label{eq:polscalar}
\end{equation}
by the induced polarized charge due to the light incidence, which originates from the electron-electron interaction.
Then, the perturbative Hamiltonian is modified as
\begin{align}
&\hat{H}^\prime \simeq - \int d\bm{x} \left[
\hat{\bm{j}} (\bm{x}) \cdot \bm{A} (\bm{x};\omega)
- \delta \hat{\rho} (\bm{x}) \Big( \phi_{\rm ext} (\bm{x};\omega) + \hat{\phi}_{\rm pol} (\bm{x}) \Big)
\right]
%\phi_{\rm ext} (\bm{x},t)
\nonumber \\
& \hspace{65mm} \times
e^{-i\omega t}.
\label{eq:Hprime2}
\end{align}

\subsection{Self-Consistent Equation}
From the Hamiltonian $\hat{H} = \hat{H}_0 + \hat{H}^\prime$, a nonlocal susceptibility is deduced.
In the formulation, we use a four-vector representation, $\bm{\mathcal{A}} = (\bm{A},-\phi/c)^{\rm t}$ and
$\hat{\bm{\mathcal{J}}} = (\hat{\bm{j}},c\delta \hat{\rho})^{\rm t}$.
%We assume a monochromatic field, $\bm{\mathcal{A}} (\bm{x},t) = \bm{\mathcal{A}} (\bm{x};\omega) e^{-i\omega t}$.
By taking the statistical average of the four-vector current,
$\langle \hat{\bm{\mathcal{J}}} (\bm{x},t) \rangle = \bm{\mathcal{J}} (\bm{x};\omega) e^{-i\omega t}$,
we obtain the constitutive equation,
\begin{equation}
\bm{\mathcal{J}} (\bm{x};\omega)
= \bm{\mathcal{J}}_0 (\bm{x};\omega)
+ \int d\bm{x}^\prime \bar{\mathcal{X}} (\bm{x},\bm{x}^\prime ;\omega)
\cdot \bm{\mathcal{A}} (\bm{x}^\prime ;\omega).
\label{eq:cons4vec}
\end{equation}
The first term is $\bm{\mathcal{J}}_0 = -(e \rho_0/m) (\bm{A},0)^{\rm t}$.
The nonlocal susceptibility is described in terms of
$\bm{\mathcal{J}}_{\nu \mu} = \langle \nu |\hat{\bm{\mathcal{J}}}|\mu \rangle$,~\cite{book:cho1}
\begin{align}
\bar{\mathcal{X}} (\bm{x},\bm{x}^\prime ;\omega)
= \sum_{\mu ,\nu}
& \left[
f_{\nu \mu} (\omega) \bm{\mathcal{J}}_{\mu \nu} (\bm{x}) \left( \bm{\mathcal{J}}_{\nu \mu} (\bm{x}^\prime) \right)^{\rm t}
\right.
\nonumber \\
& \left.
+
h_{\nu \mu} (\omega) \bm{\mathcal{J}}_{\nu \mu} (\bm{x}) \left( \bm{\mathcal{J}}_{\mu \nu} (\bm{x}^\prime) \right)^{\rm t}
\right].
\label{eq:nonlocalsus4vec}
\end{align}
Here, we consider the bases $|\mu \rangle$ and $|\nu \rangle$ from the eigenstates of $\hat{H}_0$ including the electron-electron interaction.
The factors are
$f_{\nu \mu} (\omega) = \eta_\mu/(\hbar \omega_{\nu \mu} - \hbar \omega - i\gamma)$ and
$h_{\mu \nu} (\omega) = \eta_\mu/(\hbar \omega_{\nu \mu} + \hbar \omega + i\gamma)$ with
the difference of eigenenergies $\hbar \omega_{\nu \mu} = E_\nu - E_\mu$.
The imaginary part $\gamma$ is an infinitesimal value owing to the causality or the assumed nonradiative damping constant.
The numerator is $\eta_\mu = \langle \mu |e^{-\beta \hat{H}_0}| \mu \rangle /Z_0$ with
the partition function $Z_0 = {\rm Tr} \{ e^{-\beta \hat{H}_0} \}$.

The current and charge densities $\bm{\mathcal{J}} (\bm{r};\omega)$ become the sources of response fields in the Maxwell's equations.
In the vector and scalar potential representations, the Maxwell's equations are summarized as
\begin{equation}
\bar{\mathcal{D}} (\bm{x};\omega) \bm{\mathcal{A}} (\bm{x};\omega) = - \mu_0 \bm{\mathcal{J}} (\bm{x};\omega)
\label{eq:Maxwell4vec}
\end{equation}
with
\begin{equation}
\bar{\mathcal{D}} (\bm{x};\omega) =
\left( \begin{matrix}
\bm{\nabla}^2 + (\omega/c)^2 & -i(\omega/c) \bm{\nabla} \\
0                            & - \bm{\nabla}^2
\end{matrix} \right).
\end{equation}
The formal solution of Eq.\ (\ref{eq:Maxwell4vec}) is
\begin{equation}
\bm{\mathcal{A}} (\bm{x} ;\omega) = \bm{\mathcal{A}}_0 (\bm{x} ;\omega)
- \mu_0 \int d\bm{x}^\prime \bar{\mathcal{G}} (\bm{x},\bm{x}^\prime ;\omega)
\bm{\mathcal{J}} (\bm{x}^\prime ;\omega)
\label{eq:formalA4vec}
\end{equation}
with the Green's function satisfying
\begin{equation}
\bar{\mathcal{D}} (\bm{x} ;\omega) \bar{\mathcal{G}} (\bm{x},\bm{x}^\prime ;\omega) = \delta (\bm{x} - \bm{x}^\prime).
\label{eq:Green4vec}
\end{equation}
The first term denotes an ``incident field'' as $\bar{\mathcal{D}} (\bm{x} ;\omega) \bm{\mathcal{A}}_0 (\bm{x} ;\omega) = 0$.

The constitutive equation (\ref{eq:cons4vec}) and the solution of Maxwell's equations (\ref{eq:formalA4vec}) are in a self-consistent relation.
To make it solvable form, we construct a matrix form of the self-consistent equation by multiplying
$\left( \bm{\mathcal{J}}_{\nu' \mu'} (\bm{x}) \right)^{\rm t}$, $\left( \bm{\mathcal{J}}_{\mu' \nu'} (\bm{x}) \right)^{\rm t}$, and
$\left( \varphi_{j'} (\bm{x}) \bm{e}_\beta \right)^{\rm t}$ from the left and integrating with respect to $\bm{x}$.
Here, $\varphi_{j'} (\bm{x})$ is a function for the expansion of delta function,
$\delta (\bm{x}-\bm{x}') = \sum_j \varphi_j^\ast (\bm{x}) \varphi_j (\bm{x}')$,
which is applied to $\bm{\mathcal{J}}_0 (\bm{x};\omega)$ term.
For $\left( \bm{\mathcal{J}}_{\nu' \mu'} (\bm{x}) \right)^{\rm t}$, the equation becomes
\begin{align}
\left(\hbar \omega_{\nu' \mu'} - \hbar \omega - i\gamma \right) X_{\nu' \mu'}^{(-)}
&=
Y_{\nu' \mu'}^{(0)}
+ \sum_{j,\alpha}       U_{\nu' \mu',j \alpha} X_{j\alpha}^{(A)}
\nonumber \\
& 
- \sum_{\mu,\nu} \left[ K_{\nu' \mu',\mu \nu}  X_{\nu \mu}^{(-)}
                      + L_{\nu' \mu',\nu \mu}  X_{\mu \nu}^{(+)} \right]
\label{eq:exSCmatrix1} 
\end{align}
with
\begin{align}
X_{\nu \mu}^{(-)} (\omega)
&=
\frac{1}{\hbar \omega_{\nu \mu} - \hbar \omega - i\gamma}
\int d\bm{x} \left( \bm{\mathcal{J}}_{\nu \mu} (\bm{x}) \right)^{\rm t}
\bm{\mathcal{A}} (\bm{x} ;\omega),
\label{eq:X-JA} \\
%%%
X_{\mu \nu}^{(+)} (\omega)
&=
\frac{1}{\hbar \omega_{\nu \mu} + \hbar \omega + i\gamma}
\int d\bm{x} \left( \bm{\mathcal{J}}_{\mu \nu} (\bm{x}) \right)^{\rm t}
\bm{\mathcal{A}} (\bm{x} ;\omega),
\label{eq:X+JA} \\
%%%
X_{j \alpha}^{(A)} (\omega)
&=
\int d\bm{x} \left( \varphi_j (\bm{x}) \bm{e}_\alpha \right)^{\rm t} \bm{\mathcal{A}} (\bm{x};\omega),
\label{eq:exSFXA} \\
%%%
Y^{\rm (0)}_{\nu' \mu'} (\omega)
&=
\int d\bm{x} \left( \bm{\mathcal{J}}_{\nu' \mu'} (\bm{x}) \right)^{\rm t}
\bm{\mathcal{A}}_0 (\bm{x} ;\omega)
\label{eq:Y0JA}
\end{align}
and
%%%
\begin{align}
K_{\nu' \mu' ,\mu \nu} (\omega)
&=
\mu_0 \eta_\mu \iint d\bm{x} d\bm{x}' \left( \bm{\mathcal{J}}_{\nu' \mu'} (\bm{x}) \right)^{\rm t}
\bar{\mathcal{G}} (\bm{x},\bm{x}' ;\omega) \bm{\mathcal{J}}_{\mu \nu}(\bm{x}'),
\label{eq:KJGJ} \\
%%%
L_{\nu' \mu' ,\nu \mu} (\omega)
&=
\mu_0 \eta_\mu \iint d\bm{x} d\bm{x}' \left( \bm{\mathcal{J}}_{\nu' \mu'} (\bm{x}) \right)^{\rm t}
\bar{\mathcal{G}} (\bm{x},\bm{x}' ;\omega) \bm{\mathcal{J}}_{\nu \mu}(\bm{x}'),
\label{eq:LJGJ} \\
%%%
U_{\nu' \mu' ,j \alpha} (\omega)
&=
\left( \frac{\omega_{\rm p}}{c} \right)^2
\iint d\bm{x} d\bm{x}' \left( \bm{\mathcal{J}}_{\nu' \mu'}(\bm{x}) \right)^{\rm t}
\bar{\mathcal{G}} (\bm{x},\bm{x}';\omega)
\nonumber \\
& \hspace{42mm} \times
\left( \varphi_j^\ast (\bm{x}') \bm{e}_\alpha \right).
\label{eq:exSFU}
\end{align}
$K_{\nu' \mu' ,\mu \nu}$, $L_{\nu' \mu' ,\nu \mu}$, and $U_{\nu' \mu' ,j \alpha}$ are the matrix elements in the matrix form of the self-consistent equation.
Here, $\omega_{\rm p} = \sqrt{e^2 n_0/(\epsilon_0 m^*)}$ is the plasma frequency for bulk materials with
$n_0 = (2m^* \varepsilon_{\rm F}/\hbar^2)^{3/2} / (3\pi^2)$ being the density of electrons.
The factors $X_{\nu \mu}^{(\mp)}$, $X_{j \alpha}^{(A)}$, and $Y^{\rm (0)}_{\nu' \mu'}$ are treated in vector forms as $\bm{X}$ and $\bm{Y}$.
For multiplying $\left( \bm{\mathcal{J}}_{\mu' \nu'} (\bm{x}) \right)^{\rm t}$ and $\left( \varphi_{m'} (\bm{x}) \bm{e}_\beta^{\rm t} \right)$,
we can introduce other matrix elements.
By combining the three equations, we built a matrix $\bar{\Xi} (\omega)$ in the matrix form of the self-consistent equation
\begin{equation}
\bar{\Xi} (\omega) \bm{X} (\omega) = \bm{Y} (\omega).
\label{eq:exSCmat}
\end{equation}
The matrix $\bar{\Xi} (\omega)$ includes the information of electronic excitations in a microscopic description and
the electromagnetic field acting on the excitations.
Hence, the roots of $\bar{\Xi} (\omega)$ give the spectrum of the individual and collective excitations of electrons and holes.
The detailed formulation is summarized in our previous study~\cite{Yokoyama22}.

The evaluation of the matrix elements for $\bar{\Xi} (\omega)$ is implemented by numerical calculations.
The factors $K_{\nu' \mu' ,\mu \nu}$ and $L_{\nu' \mu' ,\nu \mu}$ with the Green's function $\bar{\mathcal{G}}$
consists of the current-current, current-charge, and charge-charge interactions, which are mediated by
the T-, T-L hybridization, and L-components of the electromagnetic field in $\bar{\mathcal{G}}$.
Then, we introduce a tuning parameter $\zeta$ to switch between the presence and absence of the T field contribution in $\bar{\mathcal{G}}$.
At $\zeta = 0 (1)$, the T field is neglected (completely considered).
In the following, we take account of the effect of the background refractive index, $n_{\rm b}$.
In Eqs.\ (\ref{eq:KJGJ})--(\ref{eq:exSFU}), multiple integrals for the space coordinates are included.
By applying the Fourier transformation, the number of integrals can be reduced~\cite{Yokoyama22},
\begin{align}
K_{\mu' 0,0\mu}
&=
\mu_0 \int d\bm{k}
\left[
{\zeta^2}
\left(\tilde{\bm{j}}_{\mu' 0}(-\bm{k}) \right)^{\rm t}
\left( \frac{1}{-\bm{k}^2 + (n_{\rm b} \omega/c)^2} \right) \tilde{\bm{j}}_{0\mu}(\bm{k})
\right. \nonumber \\
& \hspace{3mm}
+ \zeta
\left(\tilde{\bm{j}}_{\mu' 0}(-\bm{k}) \right)^{\rm t}
\left( - \frac{1}{\bm{k}^2} \frac{n_{\rm b} \omega/c}{- \bm{k}^2 + (n_{\rm b} \omega/c)^2} \bm{k} \right)
(c/n_{\rm b}) \tilde{\rho}_{0 \mu}(\bm{k})
\nonumber \\
& \hspace{3mm} \left.
+
(c/n_{\rm b}) \tilde{\rho}_{\mu' 0}(-\bm{k}) \left( \frac{1}{\bm{k}^2} \right) (c/n_{\rm b}) \tilde{\rho}_{0 \mu}(\bm{k})
\right]
\label{eq:Kmu00mu}
\end{align}
Here, $n_{\rm b} \omega/c$ denotes the wavenumber of light in the nanorod and the environment with the refractive index $n_{\rm b}$.
We examine the numerical calculations for the Fourier transferred coordinates.

\begin{figure}
\centering
\includegraphics[width=90mm]{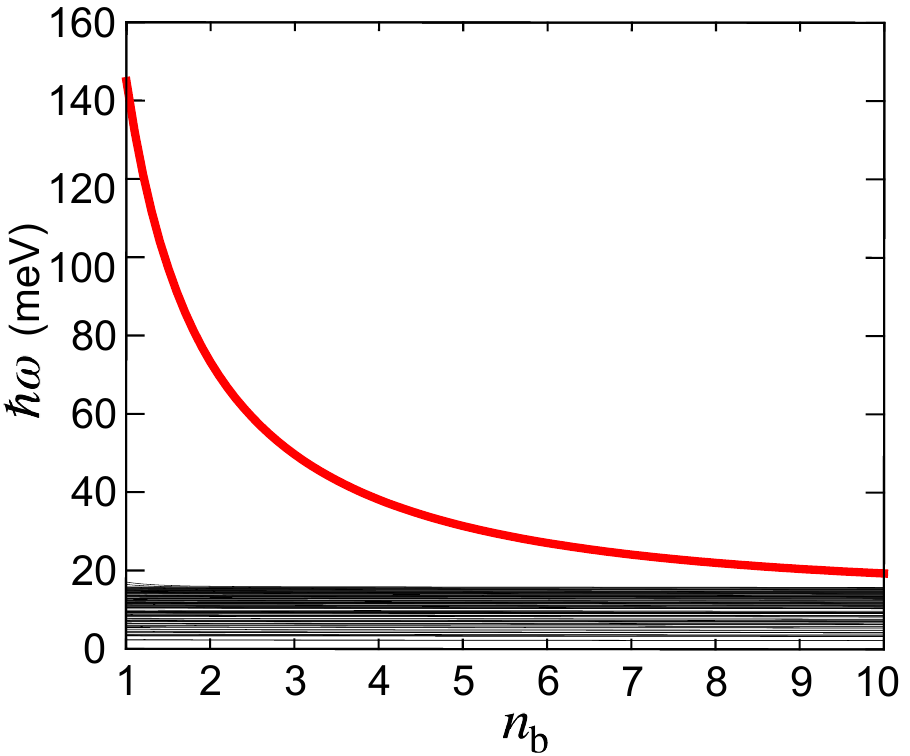}
\caption{
(Color online)
Real part of the eigenvalues of the matrix $\bar{\Xi} (\omega)$ for single rectangular nanorod with only the L field contribution ($\zeta = 0$).
The size of the nanorod is $L_x = 10\, \mathrm{nm}$, $L_y = 15\, \mathrm{nm}$, and $L_z = 500\, \mathrm{nm}$.
The Fermi energy is $\varepsilon_{\rm F} = 3\, \mathrm{eV}$.
$n_{\rm b}$ is the refractive index of the nanorod and environment.
}
\label{fig:eigenvalues}
\end{figure}

\section{Numerical Results}
\label{sec:result}
We perform numerical calculations to obtain the spectra of nanorods from the matrix $\bar{\Xi} (\omega)$.
By tuning the background dielectric constant and nanorod length, we show the contribution of the T field
and the nonlocal effect to the coherent coupling between the individual and collective excitations.

\subsection{Rectangular Nanorod}

We consider a rectangular nanorod as a plasmon-hosting nanostructure.
The electron and hole wavefunctions in a rectangular nanorod are given as
\begin{align}
&\psi_{({\rm e}\mu) = (n_x,n_y,n_z)} (\bm{x})
\nonumber \\
&= \sqrt{\frac{2}{L_x}} \sin \left( \frac{n_x \pi}{L_x} x \right)
    \sqrt{\frac{2}{L_y}} \sin \left( \frac{n_y \pi}{L_y} y \right)
    \sqrt{\frac{2}{L_z}} \sin \left( \frac{n_z \pi}{L_z} z \right)
\label{eq:wavefunc_e}
\end{align}
and
\begin{align}
&\psi_{({\rm h}\bar{\mu}) = (\bar{n}_x,\bar{n}_y,\bar{n}_z)} (\bm{x})
\nonumber \\
&= \sqrt{\frac{2}{L_x}} \sin \left( \frac{\bar{n}_x \pi}{L_x} x \right)
    \sqrt{\frac{2}{L_y}} \sin \left( \frac{\bar{n}_y \pi}{L_y} y \right)
    \sqrt{\frac{2}{L_z}} \sin \left( \frac{\bar{n}_z \pi}{L_z} z \right)
\label{eq:wavefunc_h}
\end{align}
with $L_x$, $L_y$, and $L_z$ being the lengths of nanorod in the $x$, $y$, and $z$ directions, respectively.
As a demonstration, we suppose that
$\langle \bm{x} |\mu =({\rm e} \mu ,{\rm h} \bar{\mu}) \rangle
= \psi_{({\rm e}\mu) = (n_x,n_y,n_z)} (\bm{x}) \psi_{({\rm h}\bar{\mu}) = (\bar{n}_x,\bar{n}_y,\bar{n}_z)} (\bm{x})$
is given as the basis of the Hamiltonian $\hat{H}_0$ including the electron-electron interaction
when the state $|\mu =(\rm{e}\mu ,\rm{h}\bar{\mu}) \rangle$ satisfies
$\varepsilon_{{\rm e}\mu} > \varepsilon_{\rm F} \ge \varepsilon_{{\rm h} \bar{\mu}}$.
Here, $\varepsilon_{\rm F}$ is the Fermi energy in the nanostructure.
By applying Eqs.\ (\ref{eq:wavefunc_e}) and (\ref{eq:wavefunc_h}) to the self-consistent equation,
we can discuss the effect of the spatial correlation between the T field and wavefunctions on the coherent coupling.

We calculate the eigenvalues of the system, where the complex eigenvalues are provided by the roots of ${\rm det} [\bar{\Xi} (\omega)] = 0$.
The spatial correlation between the electromagnetic field and the wavefunction of the excitations is an essential factor for the coherent coupling.
Then, we examine the modulation of the background refractive index $n_{\rm b} = \sqrt{\epsilon_{\rm b}/\epsilon_0}$ and
the nanorod length $L_z$ to tune their spatial correlation, whereas the nanorod thickness is fixed at
$L_x = 10\, \mathrm{nm}$ and $L_y = 15\, \mathrm{nm}$ to hold the subband structures by the confinement.
The Fermi energy of conduction electrons is set as $\varepsilon_{\rm F} = 3$ to $5\, \mathrm{eV}$.
The effective mass is $m^* = 0.07 m_{\rm e}$ with $m_{\rm e}$ being the electron mass in vacuum~\cite{com1}.
As a typical size scale, we use $L_0 = 100\, \mathrm{nm}$.
Then, the order of the confinement energy is $E_0 = \hbar^2 \pi^2 / (2 m^* L_0^2) \simeq 0.537\, \mathrm{meV}$.
We put $\gamma = 0.1 \times E_0$ as the nonradiative damping constant in Eqs.\ (\ref{eq:exSCmatrix1})-(\ref{eq:X+JA}).
The plasmon excitation is mainly formed by the electron excitations near the Fermi surface with
much smaller excitation energies than the plasmon energy~\cite{inaoka97jpsj}.
Hence, we focus on bare individual excitations for $(n_x, n_y, n_z)$ and $(\bar{n}_x, \bar{n}_y, \bar{n}_z) =(n_x, n_y, n_z -1)$.
This yields the smallest wavenumber in the plasmon and electron-hole pair excitation spectra,
$|\bm{q}| =\pi \sqrt{(\frac{n_x - \bar{n}_x}{L_x})^2 + (\frac{n_y - \bar{n}_y}{L_y})^2 + (\frac{n_z - \bar{n}_z}{L_z})^2} = \frac{\pi}{L_z}$.
For our consideration with $\varepsilon_{\rm F} = 3\, \mathrm{eV}$ and $L_z = 500\, \mathrm{nm}$,
the number of bases for the bare individual excitations is 114 including the spin degrees of freedom.
They are distributed in the range of $2.2\, \mathrm{meV} \le \varepsilon_{{\rm eh},\mu} \le 15.8\, \mathrm{meV}$, where
$\varepsilon_{{\rm eh},\mu} = \varepsilon_{{\rm e}\mu} - \varepsilon_{{\rm h}\bar{\mu}}$ means the excitation energy of single electron-hole pair.

%%% for reply
It is worthy to note that the three-dimensional bulk plasma frequency is evaluated as $\hbar \omega_{\rm p} \approx 2.93\, \mathrm{eV}$
when $\varepsilon_{\rm F} = 3\, \mathrm{eV}$, $m^* = 0.07 m_{\rm e}$, and $n_{\rm b} = 1$.
For the three-dimensional case, the plasma frequency at $|\bm{q}| \to 0$ is finite and constant for $|\bm{q}|$.
However, for one-dimensional nanorod we consider, the plasmon excitation also exhibits a one-dimensional-like behavior,
where the plasma frequency is proportional to $|\bm{q}|$~\cite{FriesenBergersen1980,SantoyoMussot1993}.
Hence, the evaluated collective excitation energy for the nanorod at $|\bm{q}| =\pi /L_z$ is lower than the bulk plasma frequency.
%%% for reply

%%% for reply
Here, we assume a light effective mass of electrons, such as narrow gap semiconductors.
Plasmon excitation in the narrow gap semiconductor nanostructures has been examined to extend the plasmonics to THz region~\cite{Seletskiy2011}.
Such nanostructures can be fabricated with high accuracy and the plasmon excitation is modulated by the size and shape strongly~\cite{YZhou2018}.
In the following, we consider a modulation of the collective plasmon excitation by the nanorod length $L_z$ to tune an influence of the T field.
%%% for reply

\subsection{$n_{\rm b}$- and $L_z$-Dependence of Collective Excitation}

We consider a modulation of the refractive index $n_{\rm b}$ to tune the wavelength of the electromagnetic field.
We assume that the nanorod and surrounding medium have the same $n_{\rm b}$.
Then, we avoid a cavity effect due to light reflection at the surfaces of the materials and
can focus on the effect of the pure spatial correlation of electronic wavefunctions and electromagnetic fields.

Figure \ref{fig:eigenvalues} exhibits the real parts of the eigenvalues of the matrix $\bar{\Xi} (\omega)$ in
Eq.\ (\ref{eq:exSCmat}) with only the L field contribution ($\zeta = 0$).
If the self-consistent treatment with the Maxwell's equations is not considered,
the eigenvalues are exactly equal to the electron-hole pair excitations.
In Fig.\ \ref{fig:eigenvalues}, almost all the eigenvalues are found as the individual excitations in
{$2.2\, \mathrm{meV} \le \hbar\omega \le 15.8\, \mathrm{meV}$.}
An isolated curve on the higher energy side is interpreted as the collective excitation, which is denoted as $E_{\rm c}^{\rm (L)}$.
This collective excitation is formed by the Coulomb interaction between the individual excitations.

\begin{figure}
\includegraphics[width=90mm]{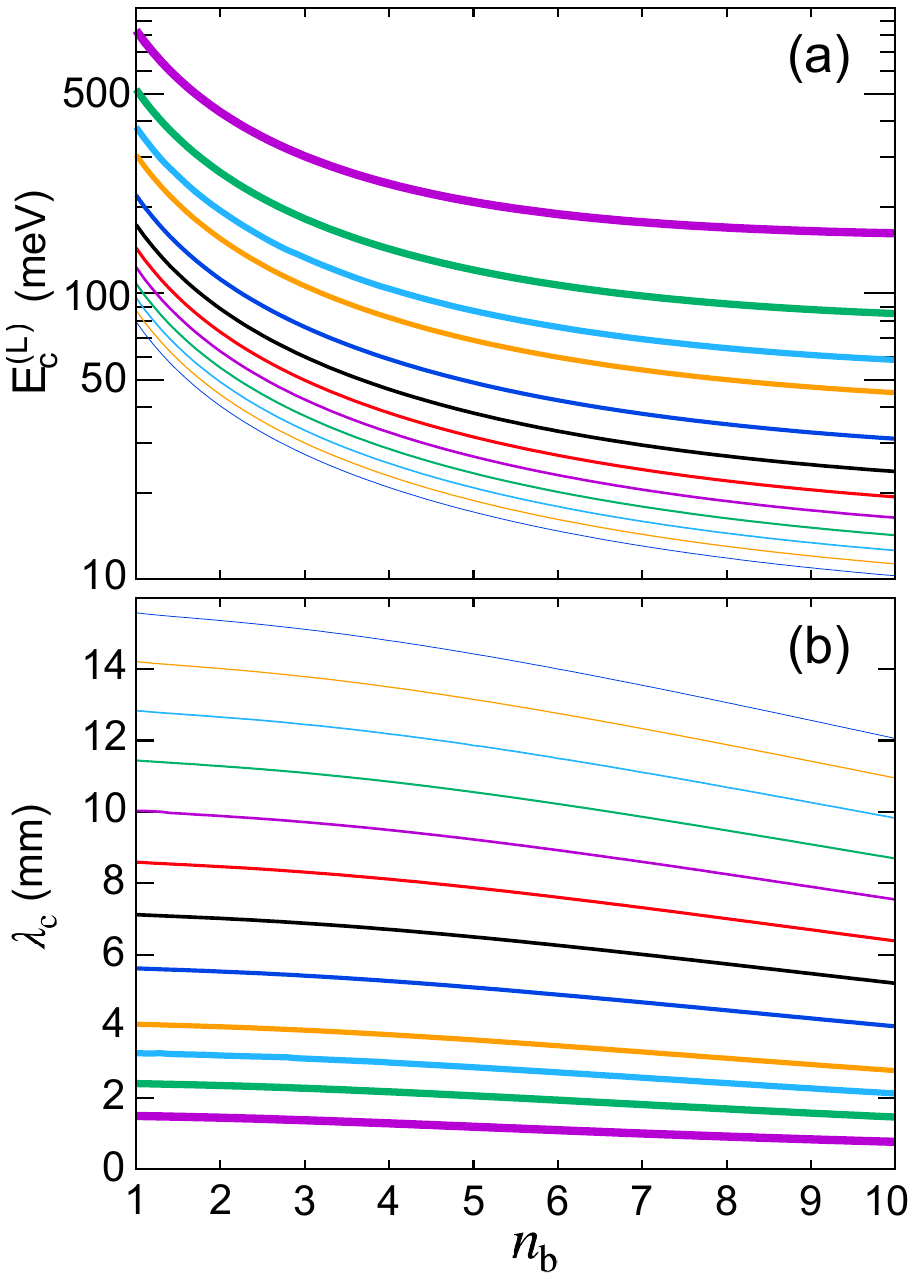}
\centering
\caption{
(Color online)
(a) Energy $E_{\rm c}^{\rm (L)}$ of collective excitation for various $L_z$
when the back ground refractive index $n_{\rm b}$ is modulated.
$L_z$ is 50 (thick line), 100, 150, 200, 300, 400, 500, 600, 700, 800, 900, and $1000\, \mathrm{nm}$ (thin line).
Red line indicates $L_z = 500\, \mathrm{nm}$ as that in Fig.\ \ref{fig:eigenvalues}.
Note that the vertical axis is log scale.
(b) Translated wavelength of (a) as $\lambda_{\rm c} = hc/(n_{\rm b} E_{\rm c}^{\rm (L)})$.
}
\label{fig:shift}
\end{figure}

The light reflection and the cavity effect can occur even by the resonant part of the refractive index.
Such effects should be significant if the material size is larger than or the same order of magnitude as the light wavelength.
However, in the present consideration, the reflection and cavity effect are not our scope and
we focus on the coherent coupling by the spatial correlation between the T field and the excitation wavefunctions.
The overall behaviors of the collective excitation against $n_{\rm b}$ seen in Fig. \ref{fig:eigenvalues}
are explained in the discussion for Fig.\ \ref{fig:shift}.

Figure \ref{fig:shift} represents $E_{\rm c}^{\rm (L)}$ for several nanorod lengths from $L_z = 50\, \mathrm{nm}$ to $1000\, \mathrm{nm}$.
Equation (\ref{eq:Kmu00mu}) reads that the effects of $n_{\rm b}$ are the modulation of the wavelength and the screening of Coulomb interaction.
When the T field is not considered, the important effect of $n_{\rm b}$ appears only through the screening.
The energy of the collective excitation, which originates from the Coulomb interaction, might be reduced monotonically by
$1/n_{\rm b}^2$ with the increase of $n_{\rm b}$.
In $L_z =500\, \mathrm{nm}$ nanorod at $n_{\rm b}=1$, the collective excitation energy is $E_{\rm c}^{\rm (L)} \simeq 144\, \mathrm{meV}$.
When $L_z =100\, \mathrm{nm}$, it is $E_{\rm c}^{\rm (L)} \simeq 518\, \mathrm{meV}$.

\begin{figure*}[t]
\includegraphics[width=170mm]{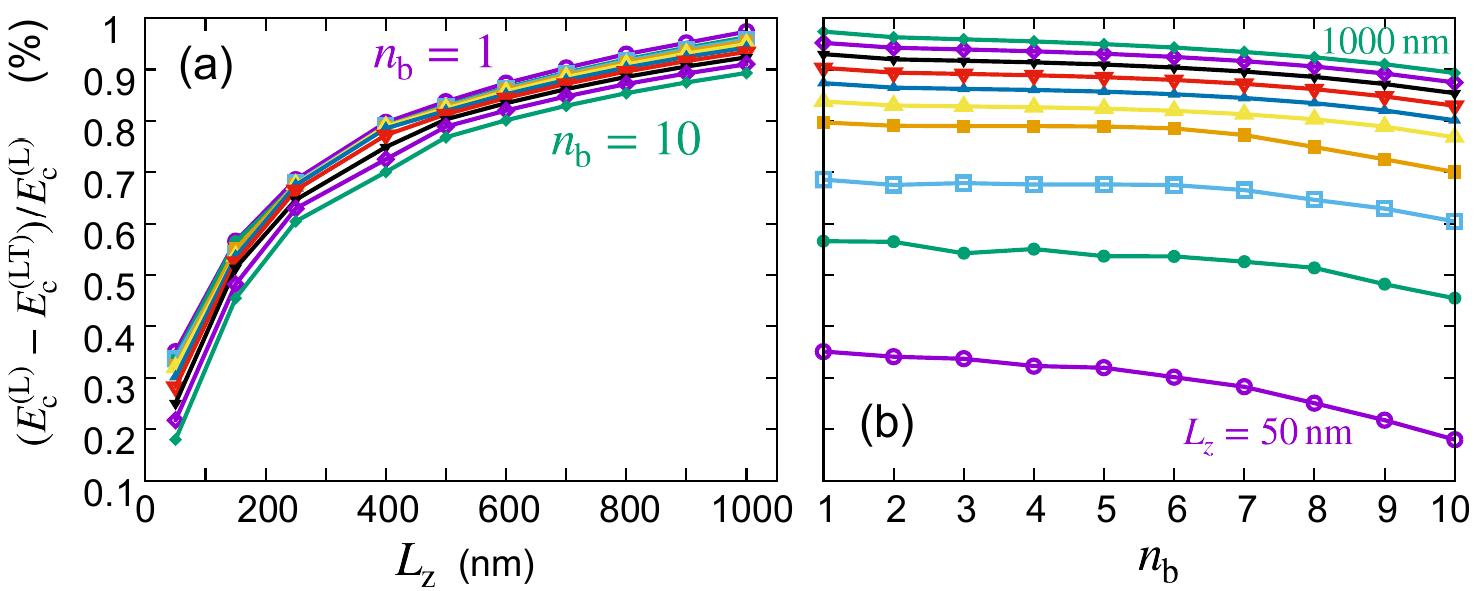}
\centering
\caption{
(Color online)
(a) Modulation ratio $(E_{\rm c}^{\rm (L)} - E_{\rm c}^{\rm (LT)})/E_{\rm c}^{\rm (L)} \times 100 \%$ as functions of $L_z$.
The plots are $n_{\rm b} = 1$ to $10$ in steps of 1.
(b) Replot of (a) as functions of $n_{\rm b}$.
}
\label{fig:ratio}
\end{figure*}

%%% for reply
As stated in section 3.1, we have considered those electronic transition processes in which
the wavenumber in the $z$ direction changes by $\pi /L_z$.
The collective excitations with $\pi /L_z$ correspond to the standing waves with the longest wavelength,
which are considered to have induced charges of opposite sign near both the nanorod ends.
Our results show that, with a decrease in $L_z$, namely, with an increase in wavenumber, the excitation energy ascends remarkably.
This is quite similar to the one-dimensional plasmon dispersion in which the energy rises up from the origin with increasing wavenumber
(see, for example, Fig.\ 6 in Ref.\ 27).
%%%

The collective excitation approaches the distributed region of the individual excitations when $n_{\rm b}$ increases.
Hence, they are indistinguishable at larger $n_{\rm b}$, which means that
the collective mode could not be formed owing to weak Coulomb interaction at large $n_{\rm b}$.
In a short nanorod, the confinement of electrons is strong.
Thus, the Coulomb interaction is effectively enlarged and the collective excitation energy is higher for a shorter $L_z$.
The above behaviors of $E_{\rm c}^{\rm (L)}$ against $n_{\rm b}$ and $L_z$ indicate that our small electronic system 
reproduces reasonable properties of collective excitations in metals.

Figure \ref{fig:shift}(b) shows the translated wavelength of the excitation energy as $\lambda_{\rm c} = hc/(n_{\rm b} E_{\rm c}^{\rm (L)})$.
The refractive index $n_{\rm b}$ shortens the wavelength.
At the same time, it suppresses the collective excitation energy $E_{\rm c}^{\rm (L)} (n_{\rm b})$ as shown in Fig.\ \ref{fig:shift}(a).
Therefore, $\lambda_{\rm c}$ does not change largely with $n_{\rm b}$.

\subsection{Effect of Transverse Field}

In the present electronic system, the T field contribution to the collective excitation energy is not prominent
if it is seen in the scale of Fig.\ \ref{fig:shift}.
Thus, to see the contribution of the T field clearly, the modulation ratio
$(E_{\rm c}^{\rm (L)} - E_{\rm c}^{\rm (LT)})/E_{\rm c}^{\rm (L)}$ is shown in Fig.\ \ref{fig:ratio}.
Here, $E_{\rm c}^{\rm (LT)}$ means the collective excitation energy when the T field contribution is fully considered ($\zeta = 1$). 
Figure \ref{fig:ratio}(a) shows $L_z$ dependence of the ratio for various values of $n_{\rm b}$.
Because the calculation including the T field is considerably heavy, we plot limited points as functions of $n_{\rm b}$.
When the nanorod length $L_z$ is longer, the distribution of individual excitations and
the collective excitation energy become lower. (No figure for the distribution of individual excitations.)
The ratio on the ordinate is always positive, and the T field contribution decreases the collective excitation energy.
As the length $L_z$ is longer, the modulation ratio is larger.
From this figure, we see that a significant T field contribution to the collective excitation appears from several hundred nanometers of $L_z$
despite the mismatch of length scale between the nanorod length ($L_z \sim 100\, \mathrm{nm}$) and the translated wavelength
($\lambda_{\rm c} = hc/(n_{\rm b} E_{\rm c}^{\rm (L)}) \sim 10000\, \mathrm{nm}$) for the plasmon ($\sim 100\, \mathrm{meV}$). 
Considering the situation that a small basis set of electronic system is used in the present calculation,
we can expect that the prominent effect of T field would appear in realistic systems.
However, we see that with an increase of $n_{\rm b}$, the modulation ratio decreases slightly for every $L_z$.
This tendency indicates that the effect of the large $n_{\rm b}$ making the light wavelength shorter is outweighed by
the effect of that reducing the collective excitation energy leading to a longer wavelength exciting the collective excitation.

We replot the data in Fig.\ \ref{fig:ratio}(a) as their dependence on $n_{\rm b}$ for the various $L_z$ as shown in Fig.\ \ref{fig:ratio}(b).
The decrease of the modulation ratio with the increase of $n_{\rm b}$ reflects
the $n_{\rm b}$ dependence of the translated wavelength $\lambda_{\rm c}$.
From this result, we understand that to enlarge the coherent coupling by the T field, 
the system size is an important parameter because of a spatial correlation between the T field and the wavefunctions of excitations.
Also, we can expect that if considering large $|\bm{q}|$ and $\varepsilon_{\rm F}$, we might obtain a situation for $L_z \sim \lambda_{\rm c}$,
where the effect by $n_{\rm b}$ modulation would be enlarged.

\subsection{Fermi Energy Dependence}

\begin{figure}
\includegraphics[width=80mm]{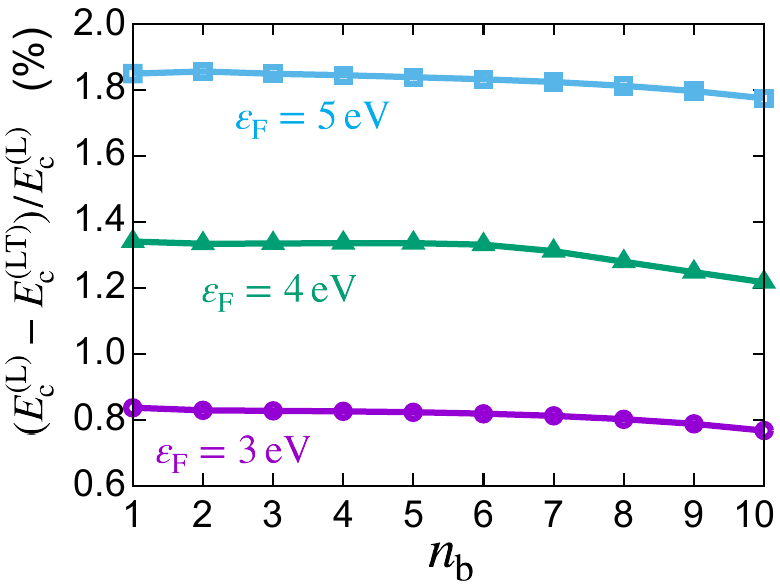}
\centering
\caption{
(Color online)
Fermi energy dependence of modulation ratio
$(E_{\rm c}^{\rm (L)} - E_{\rm c}^{\rm (LT)})/E_{\rm c}^{\rm (L)} \times 100 \%$ for $L_z = 500\, \mathrm{nm}$.
}
\label{fig:Fermi}
\end{figure}

Next, we consider the Fermi energy dependence of the T field effect.
Figure \ref{fig:Fermi} shows the modulation ratio for several Fermi energies for the electrons in the nanorod with $L_z = 500\, \mathrm{nm}$.
When $\varepsilon_{\rm F}$ is increased, the number of individual excitations contributing to form the collective excitation is increased
because the ``area'' of the Fermi surface is enlarged.
Then, the collective excitation energy becomes larger and the corresponding wavelength is shorter.
As mentioned in the explanation of Fig.\ \ref{fig:shift}, the corresponding wavelength is much longer than the rod length $L_z$ in the present situation.
Therefore, an increase of $\varepsilon_{\rm F}$ enhances the coherent coupling by the T field.
Although the modulation of $\varepsilon_{\rm F}$ is artificial in the present demonstration, 
if much larger number of the individual excitations in realistic samples is considered,
we understand that this result gives an important insight into the role of
T field-mediated coherent coupling between the collective and individual excitations.

\section{Summary and Conclusions}
\label{sec:conculusion}

Based on the microscopic nonlocal theory for metallic electronic systems, we have investigated
how the coherent coupling via the T field between the collective and individual excitations appears in the nanoscale metallic samples.
To determine the sample size at which the T-field-mediated coupling starts to become significant,
we considered a rectangular nanorod as a model sample and examined the shift of the collective excitation energy due to
the T field-mediated coupling with the individual excitations.
In the consideration, we switched the T field contribution in the Green's function representing coherent interaction between the individual excitations. 
Avoiding an overwhelming computational load, we employed the model of a small electronic basis set,
which enables us to infer the effect of an electronic system with a realistic scale from the results of a small system.
The followings are clarified from the examinations:
(1) The finite shift by the T-field mediated coupling starts to appear from several hundred nanometers of length of nanorods.
(2) Longer nanorods show a larger shift because of better spatial matching between the T field and electronic wavefunctions.
(3) A larger background dielectric constant $n_{\rm b}$ shortens the T field wavelength more.
However, simultaneously, the larger $n_{\rm b}$ reduces the collective excitation energy,
which leads to the longer wavelength exciting the collective modes.
Therefore, the larger $n_{\rm b}$ does not necessarily lead to a larger shift in the collective mode by the T field.
(4) A larger Fermi level leads to a larger energy shift due to the coherent coupling by the T field.
This is because the number of individual excitations contributing to the formation of the collective excitation is increased owing to
the enlarged ``area'' of the Fermi surface.

From the above observations, we can expect the possibility that a prominent coherent coupling 
between the collective and the individual excitations appears for a nanoscale system with sufficiently many transition paths of electrons.
These results provide guidelines to obtain a large coherent coupling between the collective and individual excitations,
which enables a bidirectional energy transfer between the photons and hot carriers.
In the next step research, we will challenge large-scale computation to handle a sufficiently large number of electrons,
and calculate the induced polarization and response field that directly reflect the quantities observed in experiments.
In particular, exploring the conditions of spontaneous resonance in the collective and individual excitations
is important for efficient creation of hot carriers with an antenna effect on the plasmons.

%\begin{acknowledgment}

\acknowledgment
The authors thank Prof.\ N.\ Yokoshi for the fruitful discussions.
This work was supported in part by JSPS KAKENHI (Grant Number: JP21H05019).

%\end{acknowledgment}

%\appendix


\begin{thebibliography}{9}
\bibitem{com1}
Assumed effective mass of electrons corresponds to the value of the conduction band of GaAs.
However, this assumption is just for a typical value of the effective mass and
our discussion is not restricted to a specific material.
\end{thebibliography}
\end{document}